\documentclass[prb,twocolumn,superscriptaddress]{revtex4}
\usepackage{graphicx}
\usepackage{dcolumn}
\usepackage{bm}
\usepackage{color}

\newcommand{\ket}[1]{\left | #1 \right \rangle}

\begin{document}

\title{Operating quantum waveguide circuits with superconducting single-photon detectors\vspace{-8pt}}

\author{C. M. Natarajan}
\affiliation{School of Engineering and Physical Sciences, Heriot-Watt University, Edinburgh, EH14 4AS UK}

\author{A. Peruzzo}
\affiliation{Centre for Quantum Photonics, H. H. Wills Physics Laboratory \& Department of Electrical and Electronic Engineering, University of Bristol, Merchant Venturers Building, Woodland Road, Bristol, BS8 1UB, UK}

\author{S. Miki}
\affiliation{National Institute of Information and Communications Technology (NICT), 4-2-1 Nukui-kitamachi, Koganei, Tokyo 184-8795, Japan}

\author{M. Sasaki}
\affiliation{National Institute of Information and Communications Technology (NICT), 4-2-1 Nukui-kitamachi, Koganei, Tokyo 184-8795, Japan}

\author{Z. Wang}
\affiliation{National Institute of Information and Communications Technology (NICT), 4-2-1 Nukui-kitamachi, Koganei, Tokyo 184-8795, Japan}

\author{B. Baek}
\affiliation{National Institute of Standards and Technology, 325 Broadway, Boulder, CO 80305, USA\vspace{-5pt}}

\author{S. Nam}
\affiliation{National Institute of Standards and Technology, 325 Broadway, Boulder, CO 80305, USA\vspace{-5pt}}

\author{R. H. Hadfield}
\affiliation{School of Engineering and Physical Sciences, Heriot-Watt University, Edinburgh, EH14 4AS UK}

\author{J. L. O'Brien}
\email{Jeremy.OBrien@bristol.ac.uk}
\affiliation{Centre for Quantum Photonics, H. H. Wills Physics Laboratory \& Department of Electrical and Electronic Engineering, University of Bristol, Merchant Venturers Building, Woodland Road, Bristol, BS8 1UB, UK}

\begin{abstract}%
Advanced quantum information science and technology (QIST) applications place exacting demands on optical components.  Quantum waveguide circuits offer a route to scalable QIST on a chip.  Superconducting  single-photon detectors (SSPDs) provide infrared single-photon sensitivity combined with low dark counts and picosecond timing resolution.  In this study we bring these two technologies together.  Using SSPDs we observe a two-photon interference visibility of $92.3\pm1.0\%$ in a silica-on-silicon waveguide directional coupler at $\lambda=804$~nm---higher than that measured with silicon detectors ($89.9\pm0.3\%$). We further operated controlled-NOT gate and quantum metrology circuits with SSPDs.  These 
demonstrations present a clear path to telecom-wavelength quantum waveguide circuits.
\end{abstract}
\maketitle

The photon \cite{loudon} is an excellent candidate for the storage and processing of quantum information \cite{nielsen}: it is well isolated from the environment even at room temperature and can be readily controlled with available optical technology.  Considerable strides have been made in the past decade in methods of generating \cite{sps}, manipulating \cite{ob-nphot-3-687} and detecting single photons \cite{ha-nphot-3}.  As a result, effects which were once curiosities of quantum optics are now exploited in fields as diverse as secure communications \cite{gi-rmp-74-145}, lithography \cite{ka-oe-15-14249}, imaging \cite{da-prl-87-013602} and metrology \cite{na-sci-316-726}; an ultimate goal is a compact device capable of scalable quantum information processing \cite{kn-nat-409-46,ob-sci-318-1567}. Meanwhile, photons remain an ideal testing ground for fundamental quantum physics and quantum information (\emph{eg.} Ref. \onlinecite{okamoto-2008}
). In this study we bring together two highly promising enabling technologies for photonic quantum information science and technology (QIST): quantum waveguide circuits and superconducting single-photon detectors (SSPDs) based on niobium nitride nanowires.

Advances in optical waveguide technology \cite{po-ieee-15-1673} can be applied to experiments on the single photon level; Quantum waveguide circuits offer a scalable route to realizing photonic QIST on a chip \cite{po-sci-320-646} (or in glass \cite{marshall-2008}):  a single silica-on-silicon waveguide chip can replace conventional bulk or fiber optical components.  In our devices the waveguide consists of a 16 $\mu$m layer of thermally grown undoped silica on a silicon wafer as the lower cladding, a 3.5 $\mu$m $\times$ 3.5~$\mu$m lithographically patterned structure of germanium and boron oxide doped silica as the core and a 16 $\mu$m phosphorous and boron doped silica layer grown atop the pattern forming the upper cladding.  These waveguide circuits support a single transverse optical mode at the design wavelength, allow evanescent coupling between adjacent waveguides and precise control of single photon states and multiphoton entanglement within a waveguide chip.   Several demonstrations have recently been carried out demonstrating the versatility and power of this technology in QIST \cite{po-sci-320-646,matthews-2008,po-sci-325-1221}.

\begin{figure}[b!]
    \centering
    \includegraphics[width = \columnwidth]{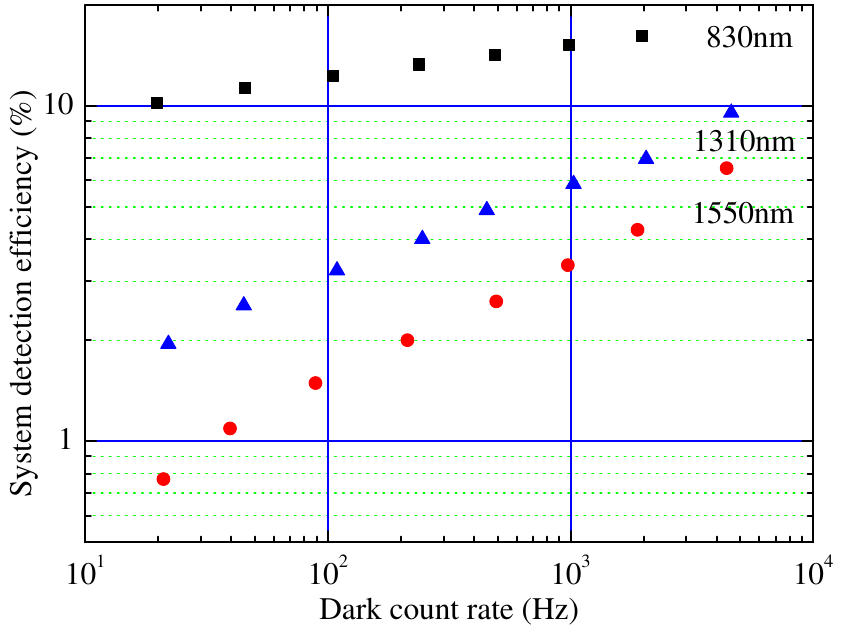}
    \vspace{-10pt}
\caption[]{\footnotesize{Superconducting single-photon detector (SSPD) system detection efficiency versus ungated dark count rate measured with calibrated attenuated laser diodes at wavelengths of 830 nm, 1310 nm and 1550 nm.  The detector operating temperature is 3 K. }}
\label{eta}
\vspace{-10pt}
\end{figure}

The demands of QIST applications have spurred the development of improved photon-counting technologies\cite{ha-nphot-3}.  Superconducting  single-photon detectors (SSPDs) \cite{go-apl-79-705} offer sensitivity from visible to mid infrared with low dark counts and excellent timing resolution.  These detectors have begun to have a significant impact on QIST applications such as QKD in optical fiber \cite{ta-nphot-1-343,tanaka-oe-16-11354}. The basic SSPD device operating principle is as follows \cite{go-apl-79-705}:  A 100 nm width wire is defined in a 4 nm thick niobium nitride film.  The wire is cooled below the superconducting transition temperature and biased close to its critical current.  When a photon strikes the wire, the current distribution is perturbed, triggering a short voltage pulse.  Our SSPD devices \cite{mi-apl-92-061116} consist of a 100 nm width meander wire defined in a 4 nm thick niobium nitride film. The devices have a 20 $\mu$m x 20 $\mu$m active area for efficient coupling to single mode telecom fiber.  The fiber-coupled SSPDs are mounted in a closed-cycle refrigerator at an operating temperature of $\sim$3 K (Ref.~\onlinecite{ha-oe-13-10846}). Our current detector system contains four SSPD channels. Figure~\ref{eta} displays the practical system detection efficiency of one of the devices used in this experiment.  Initial characterization was carried out using calibrated attenuated laser diodes at $\lambda=830, 1310$ and $1550$ nm. The full-width at half maximum (FWHM) timing jitter of the detector is 60 ps. Although our SSPDs operate at low temperatures, we use a closed cycle cryo-cooler which does not require liquid cryogens. We note that SSPDs are more robust to damage from bright light than Si single-photon avalanche diodes (SPADs).

\begin{figure}[t!]
    \centering
    \includegraphics[width = 1.0\columnwidth]{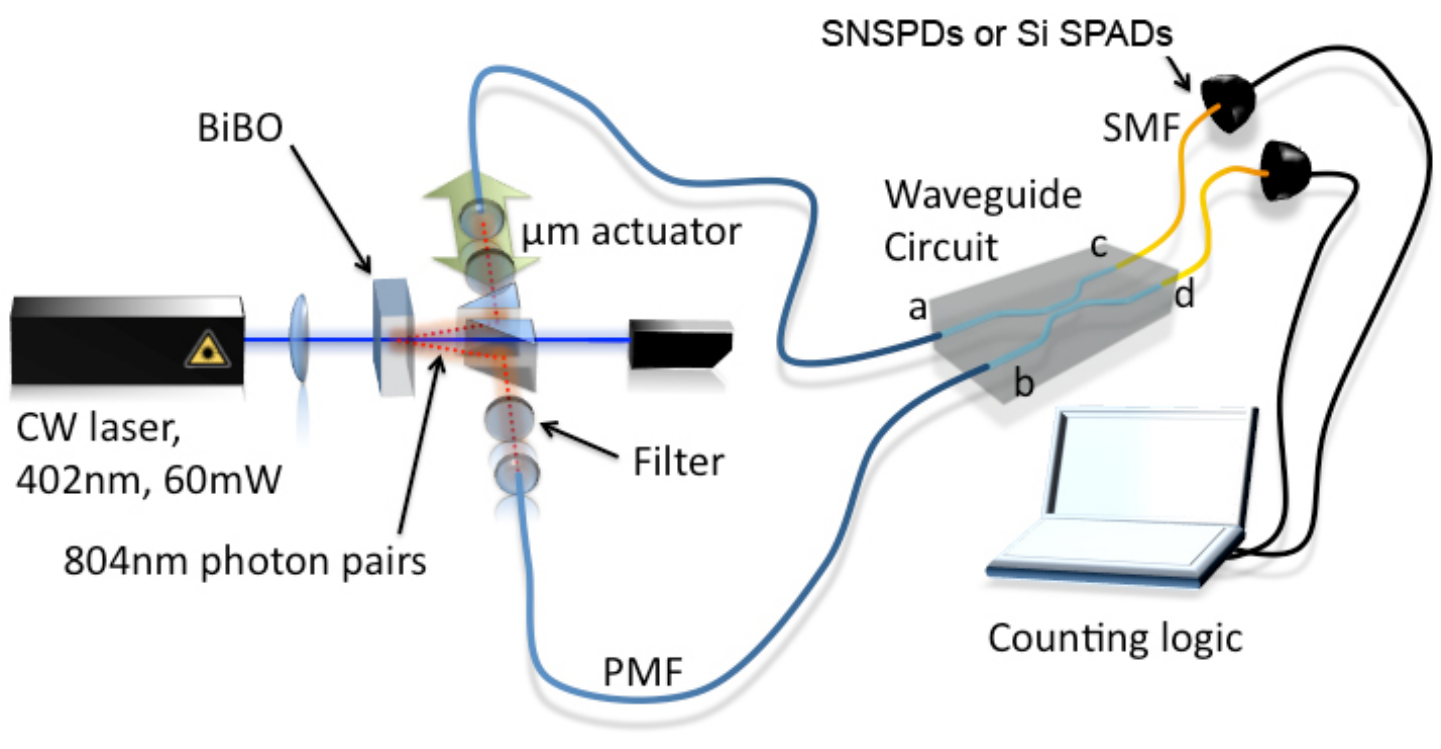}
    \vspace{-20pt}
\caption[]{\footnotesize{Experimental setup for the two-photon interference experiment.  Photon pairs at $\lambda=804$~nm are generated by spontaneous parametric down-conversion of 402 nm CW light in a Type-I nonlinear BiB$_3$O$_6$ (BiBO) crystal. Photon pairs collected and coupled to the 50:50 coupler waveguide through a polarisation maintaining fiber. The outputs of the waveguide circuit are routed to a pair of single-photon detectors (SSPDs or Si SPADs) via single mode optical fiber.  Coincidences between the detector channels are recorded using a time-correlated single-photon counting 
 card.}}
\vspace{-10pt}
\label{schem}
\end{figure}

We tested compatibility between SSPDs and quantum waveguide circuits via a two-photon interference experiment \cite{ho-prl-59-2044} (Fig.~\ref{schem}). The two-photon interference was performed using a 50:50 directional coupler waveguide circuit and detected using SSPDs. Pairs of photons at $\lambda=804$ nm were generated by spontaneous parametric downconversion of a continuous wave (CW) 402 nm laser diode `pump' beam in a type I phase matched bismuth borate (BiB$_{3}$O$_{6}$) (BiBO) nonlinear crystal crystal. The photon pair collection rate was measured as $\sim$5000 s$^{-1}$ when collected into polarization maintaining fibres (PMFs) and coupled directly to twin Si SPADs. Wavelength degenerate pairs of 804 nm photons were selected using a 2 nm bandpass filter in each path and coupled into PMFs, which were butt-coupled to the 50:50 directional coupler waveguide chip, with index matching fluid inserted at the fiber-waveguide interface.  
The output photons from the directional couplers were similarly coupled into single mode fibres (SMFs) and were detected using two channels of SSPD detector system. Overall coupling efficiencies of  70$\%$ were acheived through the waveguide (input + output insertion loss = 30$\%$). Simultaneous detection of a single photon at each output of the coupler was recorded using a time-correlated single-photon counting module with 4 ps timing resolution.  

\begin{table}[b!]
\caption[]{\footnotesize{Comparison of the properties of SSPD (measured as shown in Figure~\ref{eta}) and Si SPAD (manufacturer specifications \cite{si-apds}) detectors at $\lambda=804$ nm.}}
\label{table}
  \centering \begin{tabular}{c c c c c c}
\hline
  { } & Efficien- & Dark count  & FWHM  & $\eta^2$ &  $\eta/D\Delta t$  \\
  { } & cy $\eta$ &  rate $D$ (Hz) & Jitter $\Delta t$ (ps) &  &   \\
\hline
  Si SPAD &  0.45 & 200 & 350 & 0.203 & 6.43$\times$10$^{6}$\\
  \@804 nm\cite{si-apds} & & & & & \\
  \hline
  SSPD &  0.1 & 20 & 60 & 0.01 & 8.3$\times$10$^{7}$\\
  \@830 nm & & & & & \\
  \hline

\end{tabular}
\end{table}

Ideally, when two degenerate photons are simultaneously sent into the two input waveguides $a$ and $b$ of a 50:50 directional coupler (Fig.~\ref{schem}), quantum interference results in a path entangled state of the two photons in the two output waveguides:
\begin{equation}
\label{homeq}
\ket{1}_a\ket{1}_b\rightarrow\frac{\ket{2}_c\ket{0}_d+\ket{0}_c\ket{2}_d}{\sqrt{2}};
\end{equation}
and no simultaneous photon detection events take place due to the absence of a $\ket{1}_c\ket{1}_d$ term in this superposition \cite{ho-prl-59-2044}. In our setup, the relative arrival time of the photons at the directional coupler was varied by controlling the free space path difference using a micrometer actuator (Fig.~\ref{schem}). As this path delay is varied a `HOM' dip is observed at zero delay. 

\begin{figure}[t!]
    \centering
    \includegraphics[width = \columnwidth]{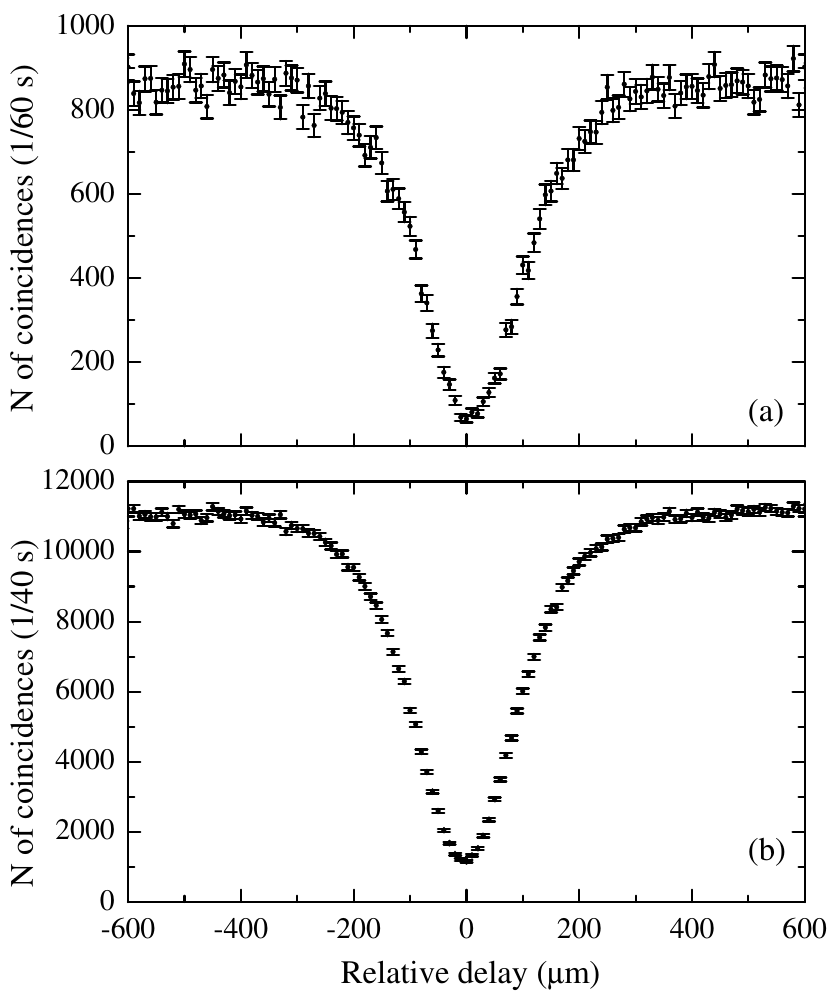}
   \vspace{-10pt}
\caption[]{\footnotesize{Two-photon interference in a quantum waveguide circuit at $\lambda=804$ nm.  HOM  dip obtained using (a) SSPDs and (b) Si SPADs.  The acquisition time per data point was 60 s for the SSPDs and 40 s for the Si SPADs.  Poissionian error bars of $\sqrt{N}$ are shown, where $N$ is the number of coincidences.
}}
\vspace{-6pt}
\label{hom}
\end{figure}

The two-photon interference experiment at $\lambda=804$ nm was performed using both SSPDs and conventional SPADs. In our experimental setup the source brightness and optical alignment remained stable over a period $\sim$1 hour, limiting the maximum duration of our experiments. The results are shown in Fig.~\ref{hom}.  The acquisition time for each data point was 60 s in the SSPD experiment (Fig.~\ref{hom}a) and 40 s in the Si SPAD experiment (Fig.~\ref{hom}b).  In both cases a high visibility ÔHOMÕ dip \cite{ho-prl-59-2044} was achieved.  However, there are noticeable differences, owing to the differing properties of the two detector types. 
The characteristics of the two detector types are given in Table I, in terms of practical detection efficiency $\eta$, ungated dark count rate $D$ and FWHM timing jitter $\Delta t$.  The accumulation rate of coincidences (off the HOM dip) was 14 s$^{-1}$ for the SSPD and 275 s$^{-1}$ for the Si SPAD (a ratio of 20:1).  This corresponds well to the square of the detection efficiency $\eta^2$ in each case (Table I).  
It is also important to consider the signal-to-noise of the two detector types.  In a time-correlated single-photon counting experiment, the effect of the dark count rate can be mitigated by gating or time binning.  The minimum effective binning interval is set by the detector jitter $\Delta t$.  A figure of merit combining these properties \cite{ha-nphot-3,RHH-FOM} and reflecting the signal-to-noise is given by $\eta/D\Delta t$.
Here, owing to the low dark counts and excellent timing jitter, the SSPD outperforms the Si SPAD by $>$10. 
We are able to exploit this advantage fully in our experimental setup, as the resolution of our timing electronics (4 ps) is well below $\Delta t$ for either detector type.  

\begin{figure}[t!]
    \centering
    \includegraphics[width = \columnwidth]{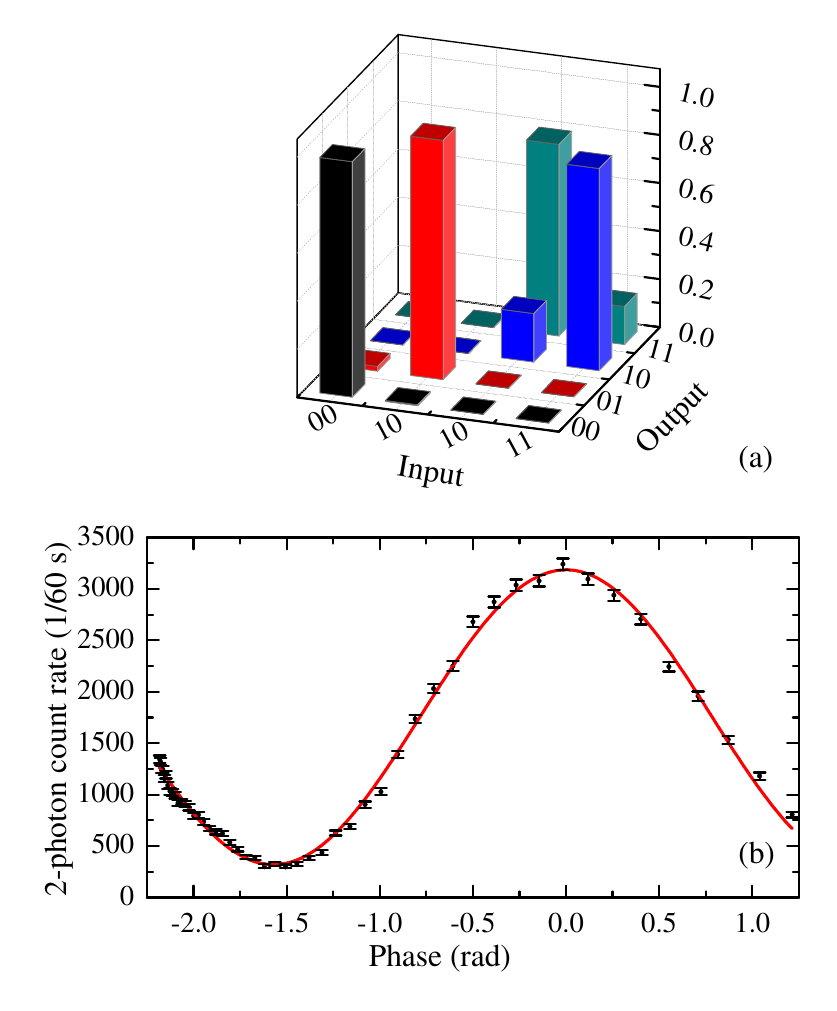}
\vspace{-20pt}
\caption[]{\footnotesize{Quantum waveguide circuits characterized with SSPDs at $\lambda=804$~nm.   (a) measured truth table for CNOT gate. (b) voltage-tuned two-photon interference in a waveguide Mach-Zehnder interferometer.}}
\vspace{-12pt}
\label{noon}
\end{figure}

This difference in signal-to-noise has a strong influence on the two-photon interference observed (Fig.~\ref{hom}) and quantified by the visibility\cite{kw-pra-45-7729} $V=(N_{max}-N_{min})/N_{max}$. 
Typically $V$ is calculated after subtracting the rate of `accidental coincidences'.  Accidental coincidences occur either due to detectors being triggered from photons in different pairs arriving within the coincidence time window, or by dark counts in one or other detector providing a spurious trigger.  The SSPD allows both contributions to be minimized due to the low timing jitter and low dark count rate.  The measured accidental coincidence rates were as $\sim$0.01~Hz for SSPDs and $\sim$5 Hz for Si SPADs.  Therefore background subtraction was unnecessary in the SSPD case. 
The raw $V$ of the HOM dip obtained using SSPDs was $92.3\pm1.0\%$, whereas that of the Si SPADs was $89.9\pm0.3\%$.  The uncertainties were calculated using the methods elaborated in Ref \onlinecite{ha-jap-101-103104}.  The larger uncertainty in the SSPD data is due to the slower coincidence accumulation rate during the overall measurement time.  The higher visibility obtained using the SSPDs is due to the better signal-to-noise (reflected by the value of the figure of merit, $\eta/D\Delta t$).  Following normal practice, the Si SPAD result can be corrected via accidental subtraction.  In this case the corrected visibility for the Si SPAD is $92.6 \pm 0.4$\%, the same as the `raw' SSPD result (non-unity visibility is attributed to the spectral distinguishability of the photons, however, it is the comparison between detector types that is important here). 

Next we used SSPDs to operate two important quantum waveguide circuits at $\lambda=804$~nm:  a controlled-NOT (CNOT) quantum logic gate comprised of 0.5 and 0.33 couplers \cite{po-sci-320-646} and a Mach-Zehnder interferometer with a voltage-controlled phase shift \cite{matthews-2008}. The CNOT gate was characterised by inputting the four computational basis states ($\ket{00},\ket{01},\ket{10},\ket{11}$) and measuring the corresponding output probabilities  (Fig.~\ref{noon}(a)); a logical basis fidelity $F=90.4\%$ was observed. 
The quantum operation of the Mach-Zehnder circuit was characterised by inputting a single photon in both inputs to generate the state of Eq.~\ref{homeq} inside the interferometer. This state exhibits an interference fringe as a function of the voltage-controlled phase that has half the period of the classical fringe (Fig.~\ref{noon}(b)); 
a contrast of  $81.8\pm 2.4\%$ was observed. The contrast and fidelity achieved show that SSPDs can be 
used to characterize advanced quantum waveguide circuits.  The contrast and fidelity is somewhat lower than that reported in Ref.~\onlinecite{po-sci-320-646, matthews-2008} using Si SPADs; this is because the SSPD results are limited by the acquisition time (determined by the stability of our current optical setup).

We have demonstrated compatibility between quantum waveguide circuits and superconducting single-photon detectors at $\lambda=804$~nm.  SSPDs offer improved signal-to-noise over Si SPAD detectors and hence give improved visibility in a HOM experiment via a 50:50 waveguide coupler. The main reason for operating these waveguides at $\lambda=804$~nm, until this point, was simply due to lack of single-photon detectors with free running operation and low dark counts at telecom wavelengths.  SSPDs present a solution to this problem \cite{ch-prl-100-133603}.   The performance of the current detectors at $\lambda=1550$~nm (Fig.~\ref{eta}.) are comparable to that of fiber-coupled devices deployed in other experiments \cite{ta-nphot-1-343}. Next generation SSPDs with improved telecom wavelength efficiency are also an imminent prospect \cite{ro-oe-14-527,mi-oe-17-23557}.  The next step is to implement quantum waveguide circuits at 1550 nm using SSPDs.  A switch to  $\lambda=1550$~nm will allow the full range of telecom waveguide technologies to be exploited 
in QIST experiments \cite{ob-nphot-3-687}.
\vspace{3pt}

\noindent {We thank Jonathan Matthews for useful discussions. This work was supported by EPSRC, QIP IRC, IARPA, ERC, the Leverhulme Trust, and NSQI. RHH is supported by a Royal Society University Research Fellowship. JLO'B acknowledges a Royal Society Wolfson Merit Award.}

%


\begin{thebibliography}{10}%
\makeatletter
\providecommand \@ifxundefined [1]{%
 \ifx #1\undefined \expandafter \@firstoftwo
 \else \expandafter \@secondoftwo
\fi
}%
\providecommand \@ifnum [1]{%
 \ifnum #1\expandafter \@firstoftwo
 \else \expandafter \@secondoftwo
\fi
}%
\providecommand \enquote [1]{``#1''}%
\providecommand \bibnamefont  [1]{#1}%
\providecommand \bibfnamefont [1]{#1}%
\providecommand \citenamefont [1]{#1}%
\providecommand\href[0]{\@sanitize\@href}%
\providecommand\@href[1]{\endgroup\@@startlink{#1}\endgroup\@@href}%
\providecommand\@@href[1]{#1\@@endlink}%
\providecommand \@sanitize [0]{\begingroup\catcode`\&12\catcode`\#12\relax}%
\@ifxundefined \pdfoutput {\@firstoftwo}{%
 \@ifnum{\z@=\pdfoutput}{\@firstoftwo}{\@secondoftwo}%
}{%
 \providecommand\@@startlink[1]{\leavevmode}%
 \providecommand\@@endlink[0]{}%
}{%
 \providecommand\@@startlink[1]{%
  \leavevmode
  \pdfstartlink
   attr{/Border[0 0 1 ]/H/I/C[0 1 1]}%
   user{/Subtype/Link/A<</Type/Action/S/URI/URI(#1)>>}%
  \relax
 }%
 \providecommand\@@endlink[0]{\pdfendlink}%
}%
\providecommand \url  [0]{\begingroup\@sanitize \@url }%
\providecommand \@url [1]{\endgroup\@href {#1}{\urlprefix}}%
\providecommand \urlprefix [0]{URL }%
\providecommand \Eprint[0]{\href }%
\@ifxundefined \urlstyle {%
  \providecommand \doi [1]{doi:\discretionary{}{}{}#1}%
}{%
  \providecommand \doi [0]{doi:\discretionary{}{}{}\begingroup
  \urlstyle{rm}\Url }%
}%
\providecommand \doibase [0]{http://dx.doi.org/}%
\providecommand \Doi[1]{\href{\doibase#1}}%
\providecommand \selectlanguage [0]{\@gobble}%
\providecommand \bibinfo [0]{\@secondoftwo}%
\providecommand \bibfield [0]{\@secondoftwo}%
\providecommand \translation [1]{[#1]}%
\providecommand \BibitemOpen[0]{}%
\providecommand \bibitemStop [0]{}%
\providecommand \bibitemNoStop [0]{.\EOS\space}%
\providecommand \EOS [0]{\spacefactor3000\relax}%
\providecommand \BibitemShut [1]{\csname bibitem#1\endcsname}%


\bibitem{loudon}%
  \BibitemOpen
  \bibfield{author}{%
  \bibinfo {author} {\bibfnamefont{R.}~\bibnamefont{Loudon}},\ }%
  \emph{\bibinfo {title} {Quantum theory of light}}\ (\bibinfo {publisher}
  {Oxford},\ \bibinfo {year} {2000})\BibitemShut{NoStop}%
\bibitem{nielsen}%
  \BibitemOpen
  \bibfield{author}{%
  \bibinfo {author} {\bibfnamefont{M.~A.}\ \bibnamefont{Nielsen}}\ and\
  \bibinfo {author} {\bibfnamefont{I.~L.}\ \bibnamefont{Chuang}},\ }%
  \emph{\bibinfo {title} {Quantum Computation and Quantum Information}}\
  (\bibinfo {publisher} {Cambridge University Press},\ \bibinfo {year}
  {2000})\BibitemShut{NoStop}%
\bibitem{sps}%
  \BibitemOpen
  \bibfield{author}{%
  \bibinfo {author} 
	 {\bibfnamefont{P.}\	\bibnamefont{Grangier}, 
	  \bibfnamefont{B.}\    \bibnamefont{Sanders}, and
	  \bibfnamefont{J.}\    \bibnamefont{Vuckovic},}
  }%
  \bibfield{journal}{%
  \bibinfo {journal} {Eds., Special Issue: Focus on Single Photons on Demand, New J. Phys.}\ }%
  \textbf{\bibinfo {volume} {6}}\ 
  (\bibinfo {year} {2004})\BibitemShut{NoStop}%
\bibitem{ob-nphot-3-687}%
  \BibitemOpen
  \bibfield{author}{%
 	\bibinfo {author} 
	 		{\bibfnamefont{J.~L.}\ \bibnamefont{O'Brien}, 
	  	 \bibfnamefont{A.}\    \bibnamefont{Furusawa}, and 
	     \bibfnamefont{J.}\    \bibnamefont{Vuckovic},}
  }%
  \bibfield{journal}{%
  \bibinfo {journal} {Nature Photon.}\ }%
  \textbf{\bibinfo {volume} {3}},\ \bibinfo {pages} {687} (\bibinfo {year}
  {2009})\BibitemShut{NoStop}%
\bibitem{ha-nphot-3}%
  \BibitemOpen
  \bibfield{author}{%
  \bibinfo {author} {\bibfnamefont{R.~H.}\ \bibnamefont{Hadfield}},\ }%
  \bibfield{journal}{%
  \bibinfo {journal} {Nature Photon.}\ }%
  \textbf{\bibinfo {volume} {3}},\ \bibinfo {pages} {696} (\bibinfo {year} {2009})\BibitemShut{NoStop}%
\bibitem{gi-rmp-74-145}%
  \BibitemOpen
  \bibfield{author}{%
  \bibinfo {author} 
	 {\bibfnamefont{N.}\	\bibnamefont{Gisin}, 
	  \bibfnamefont{G.}\    \bibnamefont{Ribordy}, 
	  \bibfnamefont{W.}\    \bibnamefont{Tittel}, and
	  \bibfnamefont{H.}\    \bibnamefont{Zbinden},}
  }%
  \bibfield{journal}{%
  \bibinfo {journal} {Rev. Mod. Phys.}\ }%
  \textbf{\bibinfo {volume} {74}},\ \bibinfo {pages} {145} (\bibinfo {year}
  {2002})\BibitemShut{NoStop}%
\bibitem{ka-oe-15-14249}%
  \BibitemOpen
  \bibfield{author}{%
	\bibinfo {author} 
	 {\bibfnamefont{Y.}\	\bibnamefont{Kawabe}, 
	  \bibfnamefont{H.}\    \bibnamefont{Fujiwara}, 
	  \bibfnamefont{R.}\    \bibnamefont{Okamoto}, 	  
	  \bibfnamefont{K.}\    \bibnamefont{Sasaki}, and
	  \bibfnamefont{S.}\    \bibnamefont{Takeuchi},}
  }%
  \bibfield{journal}{%
  \bibinfo {journal} {Opt. Express}\ }%
  \textbf{\bibinfo {volume} {15}},\ \bibinfo {pages} {14249} (\bibinfo {year}
  {2007})\BibitemShut{NoStop}%
\bibitem{da-prl-87-013602}%
  \BibitemOpen
  \bibfield{author}{%
  \bibinfo {author} 
	 {\bibfnamefont{M.}\	\bibnamefont{D'Angelo}, 
	  \bibfnamefont{Y.-H.}\    \bibnamefont{Kim}, 
	  \bibfnamefont{S.~P.}\    \bibnamefont{Kulik}, and
	  \bibfnamefont{Y.}\    \bibnamefont{Shih},}
  }%
  \bibfield{journal}{%
  \bibinfo {journal} {Phys. Rev. Lett.} \  }%
  \textbf{\bibinfo {volume} {87}},\ \bibinfo {pages} {013602} (\bibinfo {year} {2001})\BibitemShut{NoStop}%
\bibitem{na-sci-316-726}%
  \BibitemOpen
  \bibfield{author}{%
  \bibinfo {author} 
	 {\bibfnamefont{T.}\	\bibnamefont{Nagata}, 
	  \bibfnamefont{R.}\    \bibnamefont{Okamoto}, 
	  \bibfnamefont{J.~L.}\ \bibnamefont{O'Brien}, 
	  \bibfnamefont{K.}\    \bibnamefont{Sasaki}, and
	  \bibfnamefont{S.}\    \bibnamefont{Takeuchi},}
  }%
  \bibfield{journal}{%
  \bibinfo {journal} {Science}\ }%
  \textbf{\bibinfo {volume} {316}},\ \bibinfo {pages} {726} (\bibinfo {year}
  {2007})\BibitemShut{NoStop}%
\bibitem{kn-nat-409-46}%
  \BibitemOpen
  \bibfield{author}{%
  \bibinfo {author} 
	 {\bibfnamefont{E.}\	\bibnamefont{Knill}, 
	  \bibfnamefont{R.}\    \bibnamefont{Laflamme}, and
	  \bibfnamefont{G.~J.}\    \bibnamefont{Milburn},}
  }%
  \bibfield{journal}{%
  \bibinfo {journal} {Nature}\ }%
  \textbf{\bibinfo {volume} {409}},\ \bibinfo {pages} {46} (\bibinfo {year}
  {2001})\BibitemShut{NoStop}%
\bibitem{ob-sci-318-1567}%
  \BibitemOpen
  \bibfield{author}{%
  \bibinfo {author} {\bibfnamefont{J.~L.}\ \bibnamefont{O'Brien}},\ }%
  \bibfield{journal}{%
  \bibinfo {journal} {Science} \ }%
  \textbf{\bibinfo {volume} {318}},\ \bibinfo {pages} {1567} (\bibinfo {year}
  {2007})\BibitemShut{NoStop}%
  
  \bibitem[{\citenamefont{Okamoto et~al.}(2009)\citenamefont{Okamoto, O'Brien,
  Hofmann, Nagata, Sasaki, and Takeuchi}}]{okamoto-2008}
\bibinfo{author}{\bibfnamefont{R.}~\bibnamefont{Okamoto}},
  \bibinfo{author}{\bibfnamefont{J.~L.} \bibnamefont{O'Brien}},
  \bibinfo{author}{\bibfnamefont{H.~F.} \bibnamefont{Hofmann}},
  \bibinfo{author}{\bibfnamefont{T.}~\bibnamefont{Nagata}},
  \bibinfo{author}{\bibfnamefont{K.}~\bibnamefont{Sasaki}}, \bibnamefont{and}
  \bibinfo{author}{\bibfnamefont{S.}~\bibnamefont{Takeuchi}},
  \bibinfo{journal}{Science} \textbf{\bibinfo{volume}{323}},
  \bibinfo{pages}{483} (\bibinfo{year}{2009}).
  
\bibitem{po-ieee-15-1673}%
  \BibitemOpen
  \bibfield{author}{%
  \bibinfo {author} 
	{\bibfnamefont{A.}\	\bibnamefont{Politi}, 
	  \bibfnamefont{J.~C.~F.}\    \bibnamefont{Matthews},
	  \bibfnamefont{M.~G.}\    \bibnamefont{Thompson}, and 
	  \bibfnamefont{J.~L.}\    \bibnamefont{O'Brien},}
  }%
  \bibfield{journal}{%
  \bibinfo {journal} {IEEE J. Sel. Top. Quantum Electron.}\ }%
  \textbf{\bibinfo {volume} {15}},\ \bibinfo {pages} {1673} (\bibinfo {year}
  {2009})\BibitemShut{NoStop}%
\bibitem{po-sci-320-646}%
  \BibitemOpen
  \bibfield{journal}{%
    }%
  \bibfield{author}{%
  \bibinfo {author} 
	 {\bibfnamefont{A.}\	\bibnamefont{Politi}, 
	  \bibfnamefont{M.~J.}\    \bibnamefont{Cryan}, 
	  \bibfnamefont{J.~G.}\ \bibnamefont{Rarity}, 
	  \bibfnamefont{S.}\    \bibnamefont{Yu}, and
	  \bibfnamefont{J.~L.}\    \bibnamefont{O'Brien},}
  }%
  \bibfield{journal}{%
  \bibinfo {journal} {Science}\ }%
  \textbf{\bibinfo {volume} {320}},\ \bibinfo {pages} {646} (\bibinfo {year}
  {2008})\BibitemShut{NoStop}%
\bibitem{marshall-2008}%
  \BibitemOpen
  \bibfield{author}{%
  \bibinfo {author} 
	 {\bibfnamefont{G.~D.}\	\bibnamefont{Marshall}, 
	  \bibfnamefont{A.}\	\bibnamefont{Politi}, 
	  \bibfnamefont{J.~C.~F.}\    \bibnamefont{Matthews}, 
	  \bibfnamefont{P.}\ \bibnamefont{Dekker}, 
	  \bibfnamefont{M.}\    \bibnamefont{Ams}, 
	  \bibfnamefont{M.~J.}\    \bibnamefont{Withford}, and
	  \bibfnamefont{J.~L.}\    \bibnamefont{O'Brien},}
  }%
  \bibfield{journal}{%
  \bibinfo {journal} {Opt. Express}\ }%
  \textbf{\bibinfo {volume} {17}},\ \bibinfo {pages} {12546} (\bibinfo {year}
  {2009})\BibitemShut{NoStop}%
\bibitem{matthews-2008}%
  \BibitemOpen
  \bibfield{author}{%
  \bibinfo {author} 
	 {\bibfnamefont{J.~C.~F.}\    \bibnamefont{Matthews}, 
	  \bibfnamefont{A.}\	\bibnamefont{Politi}, 
	  \bibfnamefont{A.}\ \bibnamefont{Stefanov}, and 
	  \bibfnamefont{J.~L.}\    \bibnamefont{O'Brien},}
  }%
  \bibfield{journal}{%
  \bibinfo {journal} {Nature Photon.}\ }%
  \textbf{\bibinfo {volume} {3}},\ \bibinfo {pages} {346} (\bibinfo {year}
  {2009})\BibitemShut{NoStop}%
\bibitem{po-sci-325-1221}%
  \BibitemOpen
  \bibfield{author}{%
	\bibinfo {author} 
	 {\bibfnamefont{A.}\	\bibnamefont{Politi}, 
	  \bibfnamefont{J.~C.~F.}\    \bibnamefont{Matthews}, and 
	  \bibfnamefont{J.~L.}\    \bibnamefont{O'Brien},}
  }%
  \bibfield{journal}{%
  \bibinfo {journal} {Science} \ }%
  \textbf{\bibinfo {volume} {325}},\ \bibinfo {pages} {1221} (\bibinfo {year}
  {2009})\BibitemShut{NoStop}%
\bibitem{go-apl-79-705}%
  \BibitemOpen
  \bibfield{author}{%
  \bibinfo {author} 
		{\bibfnamefont{G.~N.}\	\bibnamefont{Gol'tsman}, 
		 \bibfnamefont{O.}\	\bibnamefont{Okunev},
		 \bibfnamefont{G.}\	\bibnamefont{Chulkova}, 
		 \bibfnamefont{A.}\	\bibnamefont{Lipatov}, 
		 \bibfnamefont{A.}\	\bibnamefont{Semenov}, 
		 \bibfnamefont{K.}\	\bibnamefont{Smirnov}, 
		 \bibfnamefont{B.}\	\bibnamefont{Voronov}, and 
		 \bibfnamefont{A.}\	\bibnamefont{Dzardanov},}
  }%
  \bibfield{journal}{%
  \bibinfo {journal} {Appl. Phys. Lett.}\ }%
  \textbf{\bibinfo {volume} {79}},\ \bibinfo {pages} {705} (\bibinfo {year}
  {2001})\BibitemShut{NoStop}%
\bibitem{ta-nphot-1-343}%
  \BibitemOpen
  \bibfield{author}{%
  \bibinfo {author} 
	{\bibfnamefont{H.}\	\bibnamefont{Takesue}, 
	 \bibfnamefont{S.~W.}\	\bibnamefont{Nam},
	 \bibfnamefont{Q.}\	\bibnamefont{Zhang},
	 \bibfnamefont{R.~H.}\	\bibnamefont{Hadfield},
	 \bibfnamefont{T.}\	\bibnamefont{Honjo}, 
	 \bibfnamefont{K.}\	\bibnamefont{Tamaki}, and 
	 \bibfnamefont{Y.}\	\bibnamefont{Yamamoto},}
  }%
  \bibfield{journal}{%
  \bibinfo {journal} {Nature Photon.}\ }%
  \textbf{\bibinfo {volume} {1}},\ \bibinfo {pages} {343} (\bibinfo {year}
  {2007})\BibitemShut{NoStop}%
\bibitem{tanaka-oe-16-11354}%
  \BibitemOpen
  \bibfield{author}{%
	\bibinfo {author} 
  {\bibfnamefont{A.}\	\bibnamefont{Tanaka}, 
   \bibfnamefont{M.}\	\bibnamefont{Fujiwara},
   \bibfnamefont{S.~W.}\ \bibnamefont{Nam},
   \bibfnamefont{Y.}\	\bibnamefont{Nambu}, 
   \bibfnamefont{S.}\	\bibnamefont{Takahashi}, 
   \bibfnamefont{W.}\	\bibnamefont{Maeda}, 
   \bibfnamefont{K.}\	\bibnamefont{Yoshino}, 
   \bibfnamefont{S.}\	\bibnamefont{Miki}, 
   \bibfnamefont{B.}\	\bibnamefont{Baek},
   \bibfnamefont{Z.}\	\bibnamefont{Wang},
   \bibfnamefont{A.}\	\bibnamefont{Tajima},
   \bibfnamefont{M.}\	\bibnamefont{Sasaki}, and 
   \bibfnamefont{A.}\	\bibnamefont{Tomita},}
   }%
  \bibfield{journal}{%
  \bibinfo {journal} {Opt. Express}\ }%
  \textbf{\bibinfo {volume} {16}},\ \bibinfo {pages} {11354} (\bibinfo {year}
  {2008})\BibitemShut{NoStop}%
\bibitem{mi-apl-92-061116}%
  \BibitemOpen
  \bibfield{author}{%
  \bibinfo {author} 
	{\bibfnamefont{S.}\	\bibnamefont{Miki}, 
	 \bibfnamefont{M.}\	\bibnamefont{Fujiwara}, 
	 \bibfnamefont{M.}\	\bibnamefont{Sasaki}, 
	 \bibfnamefont{B.}\	\bibnamefont{Baek}, 
	 \bibfnamefont{A.~J.}\	\bibnamefont{Miller}, 
	 \bibfnamefont{R.~H.}\	\bibnamefont{Hadfield}, 
	 \bibfnamefont{S.~W.}\	\bibnamefont{Nam}, and 
	 \bibfnamefont{Z.}\	\bibnamefont{Wang},}
  }%
  \bibfield{journal}{%
  \bibinfo {journal} {Appl. Phys. Lett.} \ }%
  \textbf{\bibinfo {volume} {92}},\ \bibinfo {eid} {061116} (\bibinfo {year}
  {2008})\BibitemShut{NoStop}%
\bibitem{ha-oe-13-10846}%
  \BibitemOpen
  \bibfield{author}{%
  \bibinfo {author} 
	{\bibfnamefont{R.~H.}\	\bibnamefont{Hadfield}, 
	 \bibfnamefont{M.~J.}\	\bibnamefont{Stevens}, 
	 \bibfnamefont{S.~S.}\	\bibnamefont{Gruber}, 
	 \bibfnamefont{A.~J.}\	\bibnamefont{Miller}, 
	 \bibfnamefont{R.~E.}\	\bibnamefont{Schwall}, 
	 \bibfnamefont{R.~P.}\	\bibnamefont{Mirin}, and 
	 \bibfnamefont{S.~W.}\	\bibnamefont{Nam},} 
  }%
  \bibfield{journal}{%
  \bibinfo {journal} {Opt. Express}\ }%
  \textbf{\bibinfo {volume} {13}},\ \bibinfo {pages} {10846} (\bibinfo {year}
  {2005})\BibitemShut{NoStop}%
\bibitem{ho-prl-59-2044}%
  \BibitemOpen
  \bibfield{author}{%
\bibinfo {author} 
	 {\bibfnamefont{C.~K.}\	\bibnamefont{Hong}, 
	  \bibfnamefont{Z.~Y.}\    \bibnamefont{Ou}, and
	  \bibfnamefont{L.}\    \bibnamefont{Mandel},}
  }%
  \bibfield{journal}{%
  \bibinfo {journal} {Phys. Rev. Lett.}\ }%
  \textbf{\bibinfo {volume} {59}},\ \bibinfo {pages} {2044} (\bibinfo {year}
  {1987})\BibitemShut{NoStop}%
\bibitem{RHH-FOM}%
  \BibitemOpen
  \bibinfo {note} {This figure of merit is widely used elsewhere in the design of quantum key distribution experiments.}
\bibitem{kw-pra-45-7729}%
  \BibitemOpen
  \bibfield{author}{%
  \bibinfo {author} 
	 {\bibfnamefont{P.~G.}\	   \bibnamefont{Kwiat}, 
	  \bibfnamefont{A.~M.}\    \bibnamefont{Steinberg}, and
	  \bibfnamefont{R.~Y.}\    \bibnamefont{Chiao},}
  }%
  \bibfield{journal}{%
  \bibinfo {journal} {Phys. Rev. A}\ }%
  \textbf{\bibinfo {volume} {45}},\ \bibinfo {pages} {7729} (\bibinfo {year}
  {1992})\BibitemShut{NoStop}%
\bibitem{ha-jap-101-103104}%
  \BibitemOpen
  \bibfield{author}{%
  \bibinfo {author} 
	{\bibfnamefont{R.~H.}\	\bibnamefont{Hadfield}, 
 		\bibfnamefont{M.~J.}\	\bibnamefont{Stevens}, 
 		\bibfnamefont{S.~W.}\	\bibnamefont{Nam}, and 
 		\bibfnamefont{R.~P.}\	\bibnamefont{Mirin},} 
  }%
  \bibfield{journal}{%
  \bibinfo {journal} {J. Appl. Phys.} \ }%
  \textbf{\bibinfo {volume} {101}},\ \bibinfo {eid} {103104} (\bibinfo {year}
  {2007})\BibitemShut{NoStop}%
\bibitem{si-apds}%
  \BibitemOpen
  \bibinfo {note} {SPCM-AQRH Data sheet},\
  \url{http://optoelectronics.perkinelmer.com/content/RelatedLinks/SpecificationSheets/SPC_PhotoDetectors.pdf}
  \BibitemShut{NoStop}%
\bibitem{ch-prl-100-133603}%
  \BibitemOpen
  \bibfield{author}{%
  \bibinfo {author} 
	{\bibfnamefont{J.}\	\bibnamefont{Chen},
	 \bibfnamefont{J.~B.}\	\bibnamefont{Altepeter},
	 \bibfnamefont{M.}\	\bibnamefont{Medic},
	 \bibfnamefont{K.~F.}\	\bibnamefont{Lee},
	 \bibfnamefont{B.}\	\bibnamefont{Gokden},
	 \bibfnamefont{R.~H.}\	\bibnamefont{Hadfield},
	 \bibfnamefont{S.~W.}\	\bibnamefont{Nam}, and 
	 \bibfnamefont{P.}\	\bibnamefont{Kumar},}
  }%
  \bibfield{journal}{%
  \bibinfo {journal} {Phys. Rev. Lett.} \ }%
  \textbf{\bibinfo {volume} {100}},\ \bibinfo {eid} {133603} (\bibinfo {year}
  {2008})\BibitemShut{NoStop}%
\bibitem{ro-oe-14-527}%
  \BibitemOpen
  \bibfield{author}{%
  \bibinfo {author} 
 {\bibfnamefont{K.~M.}\	\bibnamefont{Rosfjord}, 
  \bibfnamefont{J.~K.~W.}\	\bibnamefont{Yang}, 
  \bibfnamefont{E.~A.}\	\bibnamefont{Dauler}, 
  \bibfnamefont{A.~J.}\	\bibnamefont{Kerman}, 
  \bibfnamefont{V.}\	\bibnamefont{Anant}, 
  \bibfnamefont{B.~M.}\	\bibnamefont{Voronov}, 
  \bibfnamefont{G.~N.}\	\bibnamefont{Gol'tsman}, and 
  \bibfnamefont{K.~K.}\	\bibnamefont{Berggren},}
  }%
  \bibfield{journal}{%
  \bibinfo {journal} {Opt. Express}\ }%
  \textbf{\bibinfo {volume} {14}},\ \bibinfo {pages} {527} (\bibinfo {year}
  {2006})\BibitemShut{NoStop}%
\bibitem{mi-oe-17-23557}%
  \BibitemOpen
  \bibfield{author}{%
	\bibinfo {author} 
  {\bibfnamefont{S.}\	\bibnamefont{Miki}, 
   \bibfnamefont{M.}\	\bibnamefont{Takeda}, 
   \bibfnamefont{M.}\	\bibnamefont{Fujiwara},
   \bibfnamefont{M.}\	\bibnamefont{Sasaki}, and 
   \bibfnamefont{Z.}\	\bibnamefont{Wang},}
   }%
  \bibfield{journal}{%
  \bibinfo {journal} {Opt. Express}\ }%
  \textbf{\bibinfo {volume} {17}},\ \bibinfo {pages} {23557} (\bibinfo {year}
  {2009})\BibitemShut{NoStop}%
\end{thebibliography}
\end{document}